# Antiphase boundaries in III-V semiconductors: Atomic configurations, band structures and Fermi levels

L. Chen[1,2], L. Pedesseau[1], Y. Léger[1], N. Bertru[1], J. Even[1], C. Cornet[1*]

[1]*Univ Rennes, INSA Rennes, CNRS, Institut FOTON – UMR 6082, F-35000Rennes, France*
[2]*Université Paris-Saclay, CNRS, Laboratoire de Physique des Solides, 91405, Orsay, France*

Here, we comprehensively investigate the atomic structures and electronic properties of different antiphase boundaries in III-V semiconductors with different orientations and stoichiometries, including {110}, {100}, {111}, {112} and {113} ones, based on first-principle calculations. Especially, we demonstrate how the ladder or zigzag chemical bond configuration can lead for the different cases to a gapped semiconducting band structure, to a gapped metallic band structure or to a non-gapped metallic band structure. Besides, we evidence that the ladder APB configurations impact more significantly the Fermi energy levels than the zigzag APB configurations. We finally discuss how these different band structures can have some consequences on the operation of monolithic III-V/Si devices for photonics or energy harvesting.

## I. INTRODUCTION

Epitaxial integration of III-V semiconductors on (001) Si substrate, leveraging the benefits of both Si (e.g. earth abundance, low cost and prevalence in the electronics and PV industries) and III-V (excellent optoelectronic performances) semiconductors, raised recently a great interest for integrated photonics [1,2], solar cells [3,4] and solar water splitting [5-7]. Among the crystal defects generated at the III-V/Si hetero-interface, antiphase boundaries (APBs) which originate from polar on non-polar epitaxy [8], were considered for years as detrimental defects for devices without a clear picture on their optoelectronic contribution in the sample and the associated atomic configuration.

This atomic configuration was discussed in different works, which all reported on the relative better stability of stoichiometric APBs (same number of III-III and V-V bonds) as compared to other configurations (excess of III-III bonds or of V-V bonds) [9,10]. Nevertheless, a set of structural studies revealed the coexistence of APB lying along various different planes (and with different stoichiometries) in group III-V/group IV samples [11-16], indicating that their formation energy is not the only driving force that defines their atomic configuration. Instead, the recent theoretical and experimental clarification of the APB generation [17] and propagation [18,19] processes, confirmed that, beyond the thermodynamic stability of chemical bonds in APBs, the kinetics of atoms at the surface during or after the coalescence of III-V 3D islands contributes to define the directions along which APBs are lying, and thus their atomic configurations.

Optoelectronic properties of APBs were also investigated in few recent works. First, Tea *et. al* studied theoretically the band structure and optical properties of the GaP stoichiometric {110} APB [20]. This study was further extended to InP stoichiometric {110} APB [2], where both experimental and theoretical data were used to show that the electronic bandgap is reduced by 2D electronic localization, and that, an intrinsic and strong electron-phonon coupling arises around the stoichiometric APBs and impact the photoluminescence properties. On the other hand, the optoelectronic properties of non-stochiometric APBs was clarified only very recently [11]. In this previous work, we considered the extreme case where the non-stoichiometric {100} APBs are purely composed of either III-III bonds or V-V bonds (for GaP, GaAs and GaSb). We especially demonstrated that these APBs introduce metallic states inside the semiconductor matrix, and explained how the resulting hybrid III-V/Si structure can lead to good transport and ambipolar properties [11]. These spectacular physical properties were then used to fabricate a photo-electrode with promising performances [11]. Coherent phonon spectroscopy was also used to demonstrate the strong influence of APBs configurations on Fermi level positionning and thus on bandlineups and space charge area in GaP/Si samples [21]. Overall, these works demonstrated that the APBs should not be considered as detrimental non-radiative recombination centers, but as 2D homovalent singularities with specific symmetry properties in a bulk semiconducting matrix.

Nevertheless, optoelectronic properties were carefully determined only for extreme APBs configurations (perfectly stoichiometric {110} APB, or perfectly non-stoichiometric {100} APBs), with very different physical behaviors observed on the band structure (semiconductor vs semimetal). Especially, the origin of the bandgap opening or closing remains unclear, as well as the contribution of the atomic configuration to this process. In the present work, we comprehensively study and analyze the structural and electronic properties of different APBs based on first-principle calculations, and we show how the stoichiometry and bond configuration of these 2D homovalent singularities impact their bandstructure, in terms of bandgap opening or closing and Fermi energy levels.

## II. CLASSIFICATION OF ANTIPHASE BOUNDARIES

APBs are commonly formed during the epitaxy of Zinc-Blende III-V semiconductors on group-IV Si substrate in the diamond phase. The APBs originate from the two different ways that the III-V crystal can adopt to fit the group IV substrate orientation [1,17]. As a consequence, two crystal domains with two different phases (turned by 90° in-plane) can grow on the same substrate, separated by non-polar III-III or V-V bonds, inside the III-V matrix. Fig. 1 shows schematically the different APB structures considered in this work. For illustration purposes, the Si substrate is drawn as an atomically flat surface, as previous works experimentally shown APBs could not be related to Si steps [17,18]. For the sake of simplicity, an abrupt interface between the III-V and the Si materials is represented although compensated interfaces should be considered for a more accurate description [17,22-23]. However, we estimate that this will not affect much the electronic properties of the APB which is quite far from the III-V/Silicon interface. Fig.1a shows the typical atomic configuration of the vertical stoichiometric APBs lying along {110} planes, which have equal numbers of III-III and V-V APB bonds within the same APB, referred hereafter as {110} APBs. This is the structure also used in ref. [2] and [20]. Fig.1b shows the typical atomic configuration of the vertical non-stoichiometric APBs lying along {100} planes, which consist of pure III-III or V-V bonds within the same APB, referred hereafter as {100}-III APB and {100}-V APB, respectively. These are the structures considered in ref. [11]. All the configurations depicted in Fig. 1a or 1b correspond to APBs propagating vertically in the sample, i.e. can be obtained experimentally when the miscut is low enough or when the growth rate imbalance coefficient between the main phase and the antiphase equals to 1 [18,19].

Different tilted APBs are also considered in this work (Fig.1c-e). Fig.1c shows the typical atomic configuration of the tilted APBs lying along {111} planes, where we can find that this APB also only contains pure III-III or V-V bonds, with the same stoichiometry than the vertical non-stoichiometric {100} one. Fig.1d shows the {112} tilted APBs with III/V atomic ratio of 2/1 or 1/2. In order to better distinguish these two APBs in Fig.1d, based on the atoms forming the vertical bonds along [001] direction (i.e. the atoms forming zigzag bonds, see below), the APB on the left side is referred hereafter as {112}-V APB and the one on the right side as {112}-III APB. The tilted {113} APBs with same number of III and V APB atoms are displayed in Fig.1e. In the same way, based on the atoms forming the vertical bonds along [001] direction, the left APB is referred hereafter as {113}-V APB and the right one as {113}-III APB.

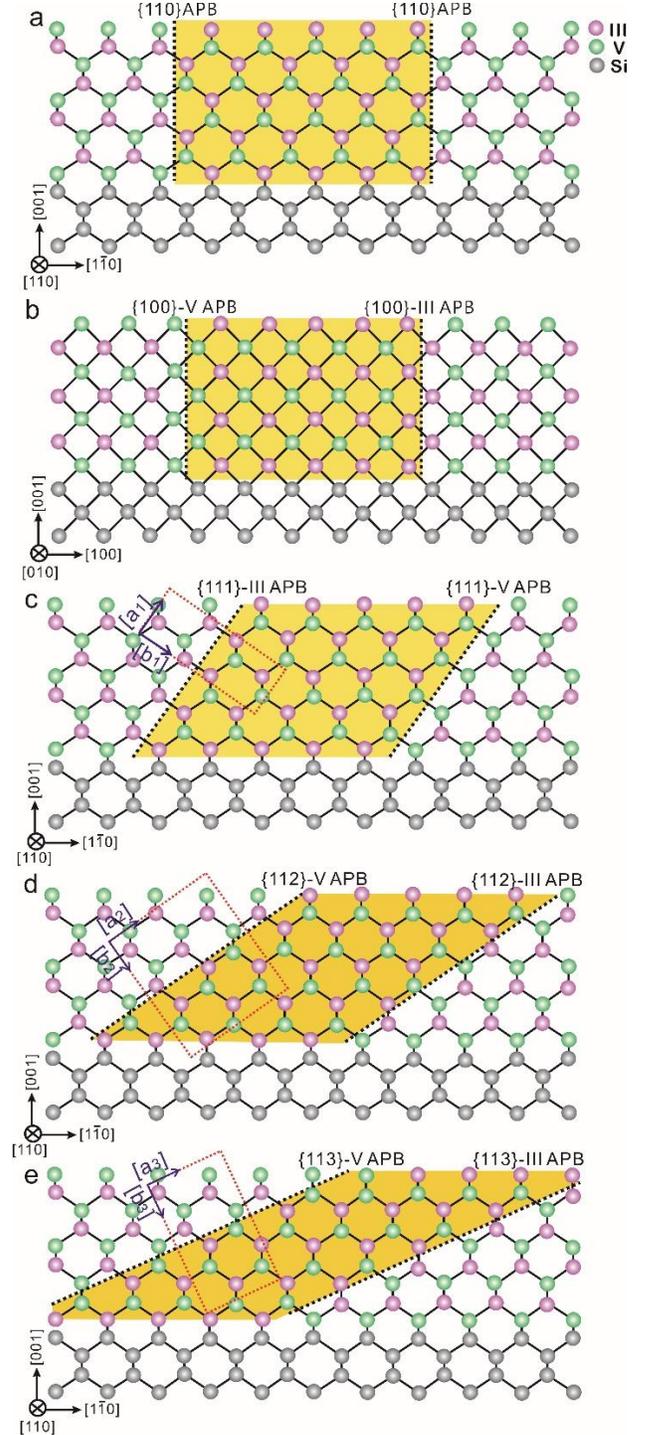

Figure 1. a-b. Schematic of vertical {110} perfectly stoichiometric (a) and {100} perfectly non-stoichiometric

(b) APBs atomic configuration in the III-V/Si (001) epitaxial materials. c-e. Schematic of the tilted {111} APB (c), {112} APB (d) and {113} APB (e) atomic configurations in the III-V/Si (001) epitaxial materials.

In order to provide a clear integrated view, the corresponding APB structures constructed for the following calculations are highlighted schematically by the red dotted boxes in the III-V/Si crystal structures (Fig.1c-e). The {111}, {112} and {113} APB planes are perpendicular to [b1], [b2], and [b3] directions, respectively, and parallel to [a1], [a2] and [a3] directions, respectively, as shown in Fig.1c-e.

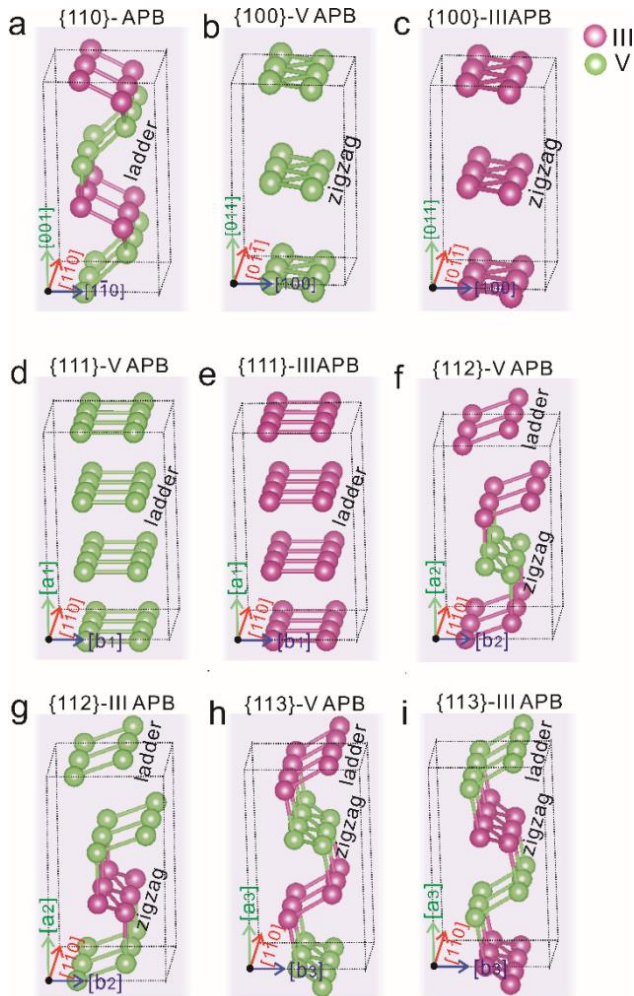

Figure 2. The APB bond configurations of the different APBs: {110} APB(a), {100}-V APB(b), {100}-III APB(c), {111}-V APB(d), {111}-III APB(e), {112}-V APB(f), {112}-III APB(g), {113}-V APB(h), {113}-III APB(i). The dotted boxes are added for a better understanding of the crystal orientations of the APB bonds.

To further investigate the atomic structures of the different APBs, the wrong bonds constituting the APB were extracted and are shown in Fig.2, where we can clearly observe that the atoms involved in the wrong bonds (referred as the "APB atoms" in this article) either form a ladder of wrong bonds, or a zigzag of wrong bonds. For the stoichiometric {110} APB, both the III-III and V-V APB atomic bonds form a ladder pattern along the [110] direction. Along the [001] direction, the III-III and V-V APB bond rows perfectly alternate with all the atoms bonded to their neighbors in the {110} planes, following a vertical snake-like arrangement (Fig.2a). For the perfectly non-stoichiometric {100} and {111} APBs, all the APB wrong bonds belong to a fixed plane family ({011} for {100} APBs and {a1} for {111} APBs). However, the structure of the APB wrong bonds rows along the [01-1] direction for {100} APBs and along the [110] for {111} APBs is completely different, as the first configuration follows a zigzag pattern (Fig.2b,c), while the second configuration follows a ladder pattern (Fig.2d,e). As compared to the {100} and {111} APBs, the {112} and {113} APBs are intermediate configurations, which contain both zigzag and ladder patterns configurations. For the {112} APB, two rows of wrong bonds following the ladder pattern and one row of wrong bonds following the zigzag pattern can be observed alternately along the [a2] direction, while for the {113} APB, a perfect alternance of ladder and zigzag patterns wrong bond rows is systematically observed along the [a3] direction.

Importantly, the relationship between the numbers of APB atoms and APB bonds are different for the zigzag and ladder APB configurations. For the zigzag bond pattern, each APB atom has two APB bonds with the neighboring APB atoms in the counter-row. Whereas, for the ladder APB bond pattern, each APB atom has one APB bond with the neighboring APB atom in the counter-row. To summarize, a number N of ladder APB atoms will lead to N/2 APB bonds, while a number N of zigzag APB atoms will lead to N APB bonds (as shown in the schematic plot of Fig.S1). Therefore, the atomic stoichiometry of the APBs (i.e. III/V APB atom ratios) are not equal to the corresponding III-III/V-V APB bond ratios for the {112} and {113} APBs due to their mixture of the different zigzag and ladder APB bond patterns. This also impacts the excess/lack of charge carried by the APB, as will be discussed later.

## II. COMPUTATIONAL METHOD

In this work, first-principles calculations of different APB structures were performed by using the Vienna ab initio simulation package [24,25] within the density functional theory (DFT) [26] applied on the binary InP material. We point out, from our previous works on APBs band structures, that similar APBs have apparently similar impact on the band structures of the various III-V semiconductors [11]. In order to investigate separately the APB singularities, slabs with specific APB in the middle of the structures were constructed by adding vacuum on both truncated sides (left top and right). Each truncated surface has been passivated with fictitious hydrogen (as in previous study [17,27]) atoms to avoid localized surface states, as shown in Fig.3. $1s^{0.75}$, and $1s^{1.25}$ were used as valence electrons for the fictitious H* with a net charge of 0.75e to compensate the V element, and the fictitious H* with a net charge of 1.25e to compensate the III element. The thicknesses of the vacuum regions are large (around 20Å and 30 Å vacuum regions for the relatively short and long slab structures) to reduce interaction between the slab surfaces. The standard generalized gradient approximation (GGA) parameterized by Perdew-Burke-Ernzerhof (PBE) [28] was used for structure optimization and the structures were relaxed until the Hellmann-Feynman forces on the atoms are less than $10^{-4}$ eV/Å. The energy cutoff was set to 500 eV.

As the common LDA and GGA calculations underestimate the electronic band gaps of the III-V semiconductors, the band structure calculations were performed with the projector augmented-wave method [29] based on expensive Heyd–Scuseria–Ernzerhof (HSE) hybrid functionals [30,31]. The accuracy of HSE band structure calculation was firstly confirmed by the calculation result of InP zinc blende structure with 1.34eV bandgap (Fig.S2), which shows good agreement with the experimental data [32]. The 3D band structures (with $k_z=0$, the z axis of the reciprocal space corresponding to the crystallographic directions perpendicular to the different APB planes) and total density of states (DOS) calculations with dense k points (ka>420Å, the length of each lattice vector (a) multiplied by the number of k points in this direction (k) is larger than 420 Å) were performed by generating ab initio tight-binding Hamiltonians from the maximally localized Wannier functions within HSE functional [33], as implemented in Wannier Tools package [34]. Wannier Tools package [35-38] has mature modules to extract 3D band structure and topology-related properties of different materials [39-42]. For this study, the Bloch wave functions were projected onto the s, p, d atomic orbitals of In and s, p atomic orbitals of P. Besides, DFT calculations of long structures were carried out based on metaGGA (TB-mBJ) [43] functionals (which are relatively low time-consuming but provides similar accuracy for the electronic band gap as compared to the expensive HSE or GW method) to investigate the APB length effect and further analyze the Fermi energy levels.

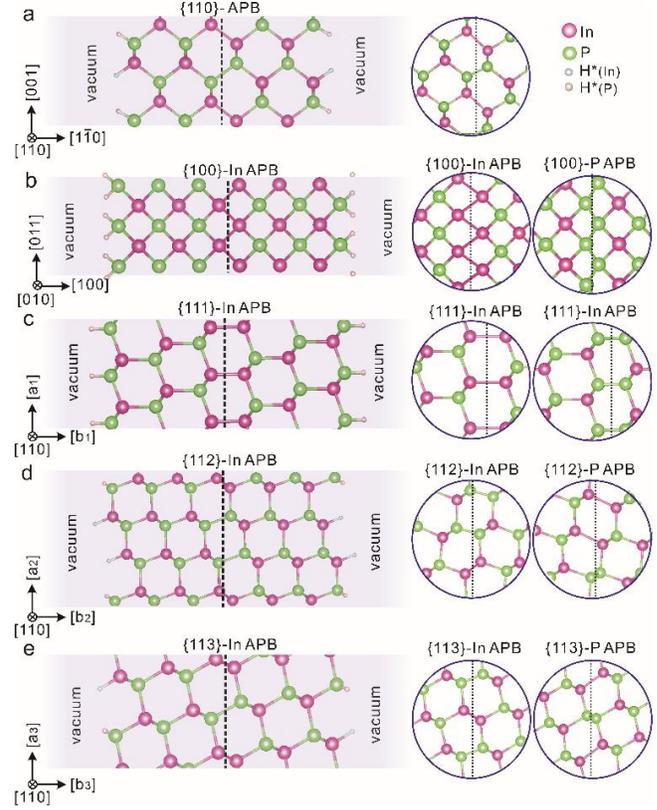

Figure 3. Different APB slab structures for theoretical calculations: {110} APB(a), {100} APB (b), {111} APB (c), {112} APB (d), {113} APB (e). The complete slab structures of {100}- P, {111}-P, {112}-P, {113}-P APBs aren't shown here directly, but they can be obtained based on the corresponding -In APBs structures (b-e) just by exchanging the In and P atoms. The small sectional images of APB structures on the right side correspond to the structures after relaxation.

## IV. RESULTS AND DISCUSSIONS

Figure 3 shows the different APB slab structures for theoretical calculations: {110} APB belonging to the Pnnm space group (a), {100} APB belonging to the Pmma space group (b), {111} APB belonging to the P-3m1 space group (c), {112} APB belonging to the P21/m space group (d), {113} APB belonging to the C2/m space group (e). All the APB structures present an inversion symmetry, but the atomic motifs are different. The complete slab structures of {100}-P, {111}-P, {112}-P, {113}-P APBs are not shown here, which can be obtained by inversion of the group-III and group-V atoms based on the corresponding -In APBs (Fig.3b-e). The small sectional images of APB structures on the right side correspond to the structures after relaxation, where we can observe that all III-III APB bonds lengthen and V-V APB bonds shorten during the relaxation.

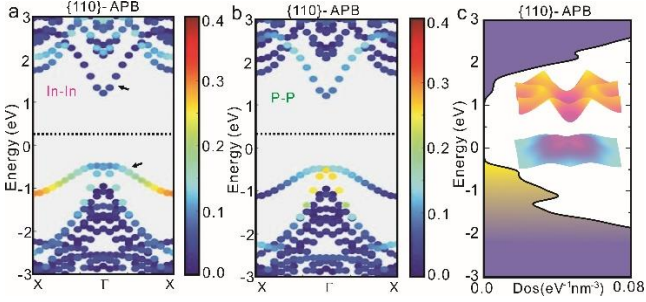

Figure 4. **Band structure and DOS calculations of vertical {110} APB**. a,b. Atom-projected band structures of {110} APB on III-III APB atoms (a) and V-V APB atoms (b). The color bar from blue to red underlines the increase of localization effect of the states at the APB atoms. c. 3D band structures at $k_z=0$ plane and corresponding calculated density of states. The inset shows the 3D band structure corresponding to the main APB quasi-localized states (on the top of the valence band) and the first conduction band marked by the arrows in (a).

Then the electronic bands are studied for the different APBs configurations studied here. Firstly, for the stoichiometric {110} APB with a ladder pattern wrong bond configuration (Fig.2a), the band structures are calculated along the following k-path X(-0.5,0,0)- Γ(0,0,0)-X(0.5,0,0) at the HSE functional level of theory, where the x axis of the reciprocal space corresponds to the [110] crystallographic direction (Fig.3a). Note that the common k-paths chosen in all the cases for the sake of comparison lay in the APB plane to illustrate the effects of the 2D electronic dispersion. Besides, the k-paths used for band structure calculations for all the APB structures are systematically chosen along the directions corresponding to the rows of similar wrong bonds (i.e. [110] crystallographic direction for {110}, {111}, {112}, {113} APBs, and [01$\bar{1}$] crystallographic direction for {100} APBs). For the 3D band structures, we focus on the $k_{x,y}$ plane with $k_z=0$, laying in the APB plane correspondingly.

To better specify the contribution of the APB atoms to the electronic states, APB atom-projected band structures on In-In APB atoms (Fig.4a) and P-P APB atoms (Fig.4b) are extracted, respectively, where the color scale indicates the spatial localization of the electronic state in the APB plane. The color map from blue to red underlines the increase of localization effect of the states at the APB atoms. We can observe that the band structure exhibits larger bandgap than the bulk Zinc-Blende InP band structure (Fig.S2) due to quantum confinement effect. In addition, band structures of the APB have more bands that result of a band folding effect compared to the bulk Zinc-Blende InP. More importantly, it can be observed that the APB introduces two localized states (LS) related respectively to III-III ladder bonds and V-V ladder bonds, at the top of the valence band (VB), shifting the valence band maximum (VBM) upward and thus reducing the electronic band gap. In order to provide a comprehensive view on the electronic properties of the APB, 3D band structures of the first CB and the APB quasi-localized band on the top of the VBs (marked by the arrows in Fig.4a) at $k_z=0$ plane, as well as the DOS calculation in the whole reciprocal space are extracted (see Fig.4c). This confirms once again the semiconducting nature of this kind of APB and shows good agreement with the results of the literature [2,20]. Specifically, this modified gapped semiconducting band structure is classified in this work as a Type-A band structure.

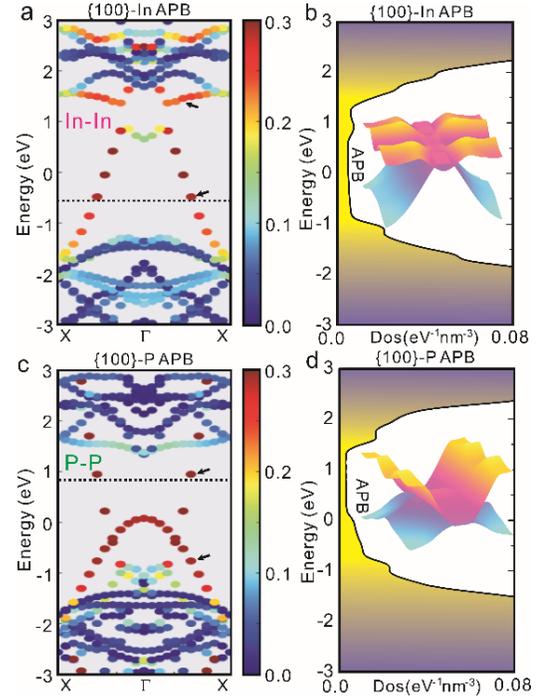

Figure 5. **Band structure and DOS calculations of vertical {100}-In (a,b) and {100}-P (c,d) APBs**. a,c. Atom-projected band structures of {100}-In APB on III-III APB atoms (a) and {100}-P APB on V-V APB atoms (c). b,d. 3D band structures (at $k_z=0$ plane) and density of states. The inset shows the 3D band structure for {100}-In (b) and {100}-P APB(d) corresponding to the main bands localized around the APB, which are marked by the arrows in (a) and (c).

Then, the electronic properties of the vertical perfectly non-stoichiometric {100} APBs with the zigzag pattern of the wrong bond configuration (Fig. 2b,c) are investigated. The APB atom-projected band structure of {100}-In APB and {100}-P APB are shown in Figure 5a and c, respectively. We can find that both the {100}-In APB and {100}-P APB structures introduce metallic states (MS) in the electronic band gap (in dark red in Fig. 5a and c), connecting the bulk-like VBs and CBs and as a consequence closing the bandgap. The metallic states are specifically localized on the APB atoms. This is further confirmed by the 3D band structures of APB localized states (marked by the arrows in Fig.5a and 5c) and DOS results, which reveal the absence of electronic band gap (Fig.5b and d). This is consistent with our previous study in ref. [11]. The main APB metallic states of {100}-In and -P APB show roughly pointed-parabola and inverted pointed-parabola along the X-Γ-X direction, lying in the bulk band gap, respectively. Besides, the Fermi energy level of the In-In rich {100}-In APB is low and close to bulk-like VBs states, while the Fermi energy level of the P-P rich {100}-P APB is high and close to bulk-like CBs states, which indicates they could behave similarly to p-doped and n-doped semiconductors [44,45], respectively. This non-gapped metallic band structure is referred hereafter as Type-B band structure.

Interestingly, these results unambiguously show that even though both {110} and {100} APBs are composed of III-III and/or V-V bonds, their electronic band properties are drastically different in terms of Fermi energy level positioning, band structure shape and also electronic band gap. In the following, we will explore the impacts of various stoichiometries and bond configurations by considering different APBs with different structures.

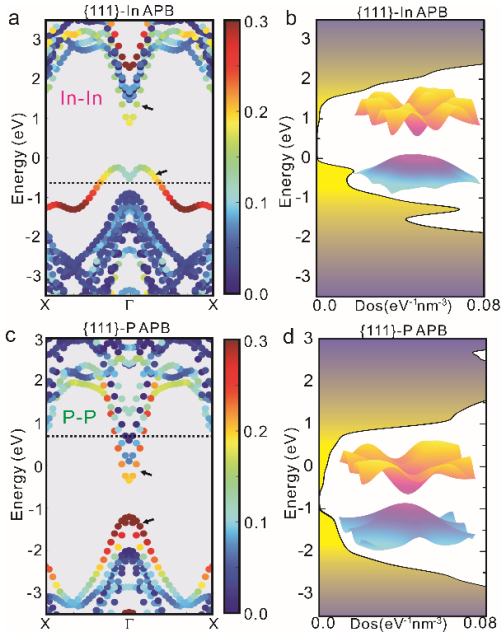

Figure 6. **Band structure and DOS calculations of vertical {111}-In (a,b) and {111}-P (c,d) APBs.** a,c. Atom-projected band structures of {111}-In APB on III-III APB atoms (a) and {111}-P APB on V-V APB atoms (c). b,d. Calculated density of states. The inset shows the 3D band structures for {111}-In (b) and {111}-P (d) APB corresponding to the main localized APB states, which are marked by the arrows in (a),(c).

The electronic band structures of the tilted and perfectly non-stoichiometric {111} APB with a ladder pattern wrong bond configuration (Fig. 2d,e) are shown in Fig.6. The atom-projected band structure of {111}-In APB on III-III APB atoms and {111}-P APB on V-V APB atoms are extracted and shown in Figure 6a and c, respectively. Interestingly, another type of band structure is obtained. Firstly, compared to the bulk material, it can be observed that these APB structures introduce localized APB states (LS) on both the top of the bulk-like VBs and bottom of the bulk-like CBs (mainly from green to dark red in Fig. 6a and c), thus reducing the band gap. But the band gap isn't closed entirely, as in the case of the {110} APB, which is further confirmed by the 3D band plots of these new localized APB bands and the DOS calculation performed for the whole reciprocal space (Fig. 6 b,d ). However, in this case, the Fermi energy level for the {111}-In APB is quite low and crosses the localized APB states on the top of the VBs, while the Fermi energy level for the {111}-P APB is high and crosses the localized APB states on the bottom of the CBs, indicating these two localized APB states are then metallic states, which could also behave as highly p-doped and n-doped semiconductors [44,45] similarly to {100} APBs. This gapped metallic APB band structure is called hereafter as Type-C band structure.

Similarly, the band structures of the tilted {112} APBs with both ladder and zigzag patterns wrong bond configurations have been also studied. Fig. 7a and b show band structures of {112}-In APB projected on III-III APB atoms with a zigzag pattern wrong bond configuration (a) and V-V APB atoms with a ladder pattern wrong bond configuration (b), respectively. Fig.7 d and e show band structures of {112}-P APB projected on III-III APB atoms with a ladder pattern wrong bond configuration (d) and V-V APB atoms with a zigzag pattern wrong bond configuration (e), respectively. Firstly, it can be observed that both {112}-In and {112}-P APB structures introduce two kinds of new states localized around APB atoms, one on the top of the VBs and the other spanning over the whole bandgap. This is further confirmed by the 3D bands structures and DOS results (Fig. 7 c, f). More specifically, we can find that for the {112}-In APB, the V-V APB bonds following a ladder pattern mainly contribute to the new localized state on the top of the VBs (in red/orange in Fig. 7b), while the metallic states connecting the CBs and VBs with a pointed-parabola shape is localized on the III-III APB atoms following a zigzag pattern (in dark red in Fig. 7a). Similarly, for the {112}-V APB, the new localized state on the top of the VBs is localized around the III-III APB atoms following a ladder pattern (in red/orange in Fig. 7d) and the metallic state connecting the CB and the VB (in dark red in Fig. 7e), with inverted pointed-parabola is localized on the V-V APB atoms following a zigzag pattern. This modified band structure is thus a combination of Type-A and Type-B band structures. Besides, the Fermi energy level of the {112}-In APB is higher than the one of the {112}-P APB suggesting a possible link between the atomic stoichiometry and the Fermi level positioning, which we will see is not that simple.

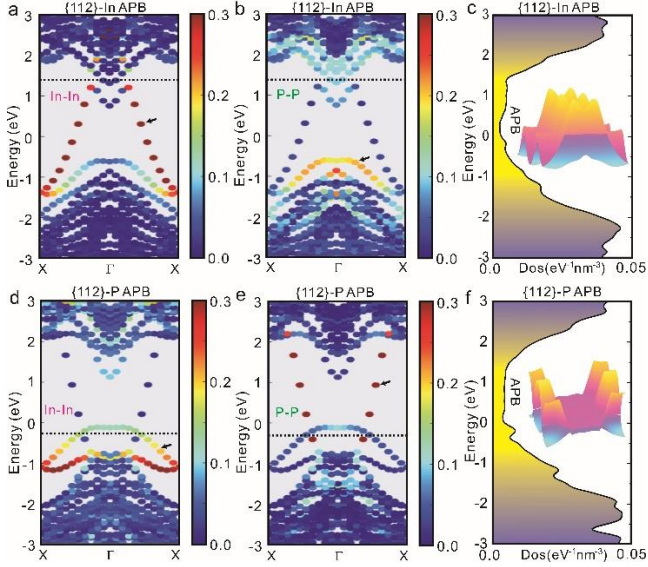

Figure 7. **Band structure and DOS calculations of {112}-In (a-c) and {112}-P (d-f) APB.** a,b. Atom-projected band structures of {112}-In APB on III-III APB atoms(a) and V-V APB atoms(b). d,e. Atom-projected band structures of {112}-P APB on III-III APB atoms(d) and V-V APB atoms(e). c,f. Calculated density of states for the {112}-In APB (c) and {112}-P APB (f) structures. The inset shows the 3D band structures corresponding to the main localized APB bands, which are marked by the arrows in (a),(b) for {112}-In APB and in (d),(e) for {112}-P APB.

Then, the band structure calculations were also performed for the tilted and atomically stoichiometric {113} APBs with mixed ladder and zigzag patterns wrong bond configurations. Fig. 8a and b show the band structures of {113}-In APB projected on III-III APB atoms following a zigzag pattern wrong bond configuration (a) and V-V APB atoms following a ladder pattern wrong bond configuration (b), respectively, and Fig.8 d and e show band structures of {113}-P APB projected on III-III APB atoms following a ladder pattern wrong bond configuration (d) and V-V APB atoms following a zigzag pattern wrong bond configuration (e), respectively. Again, we can find that both {113}-In and {113}-P APB structures introduce two kinds of new states localized around the APB, one on the top of the VBs and the other spanning over the whole bandgap. This is further confirmed by the 3D band structures and the DOS results (Fig. 8 c and f). Precisely, it can be observed that for the {113}-III APB, the V-V APB atoms following a ladder pattern wrong bond configuration give main contribution to the new localized state on the top of the VBs (mainly from green to red in Fig. 8b), while the metallic state connecting the CBs and VBs with pointed-parabola shape is localized in the III-III APB atoms following a zigzag pattern wrong bond configuration (in dark red in Fig. 8a). For the {113}-V APB, the new localized state on the top of the VB is localized on the III-III APB atoms following a ladder pattern wrong bond configuration (mainly from green to red in Fig. 8d) and the metallic state in the whole gap with inverted pointed-parabola is localized on the V-V APB atoms following a zigzag pattern wrong bond configuration (in dark red in Fig. 8e). As expected, this modified band structure is also a combination of Type-A and Type-B bands due to the mixture of ladder and zigzag pattern wrong bonds configurations. Besides, even though the {113} APBs are atomically stoichiometric, the Fermi energy level of {113}-In APB and {113}-P APB are quite different: {113}-In APBs have a relatively high Fermi energy level while {113}-P APBs have a relatively low Fermi energy level, which indicates that the atomic stoichiometry is not the pertinent determining factor governing the positioning of the Fermi energy level.

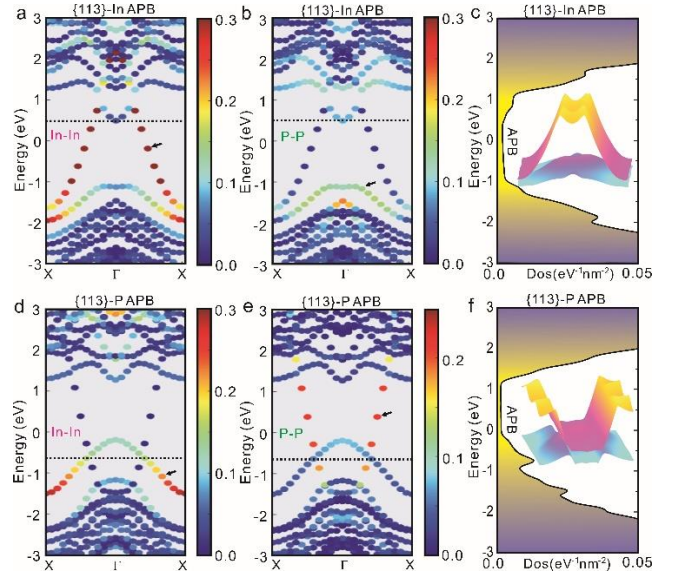

Figure 8. **Band structure and DOS calculations of {113}-In(a-c) and {113}-P(d-f) APB.** a,b. Atom-projected band structures of {113}-In APB on III-III APB atoms(a) and V-V APB atoms(b). d,e. Atom-projected band structures of {113}-P APB on III-III APB atoms(d) and V-V APB atoms(e).c,f. Calculated density of states for the {113}-In APB (c) and {113}-P APB (f) structures. The insets show the 3D band structures corresponding to the main localized APB bands, which are marked by the arrows in (a),(b) for {113}-In APB and in (d),(e) for {113}-P APB.

Combining all the band structure results obtained for the {110}, {100}, {111}, {112} and {113} APBs, we can draw the conclusions below: (1) all the APB structures introduce new states localized around APB atoms. (2) the APB structures with APB atoms only following the ladder pattern configuration will lead to localized states on the top of VBs and/or at the bottom of CBs. (3) the APB structures with APB atoms only following the zigzag pattern configuration will lead to metallic states localized in APBs, connecting the bulk-like VBs and CBs and closing the electronic band gap. More specifically, the III-III APB atoms following the zigzag pattern wrong bond configuration will lead to a pointed-parabola shape band and V-V APB atoms following the zigzag pattern wrong bond configuration will lead to an inverted pointed-parabola shape band along the X-Γ-X direction. (4) the intermediate APB structures constituted of APB atoms following both the ladder and zigzag patterns wrong bond configurations will lead to mixed bands with following the properties described in both cases (2) and (3). (5) atomic stoichiometry cannot explain alone the positioning of the Fermi energy level. An in-depth analysis is thus needed to understand the origin of the observed Fermi levels variations. This is discussed hereafter.

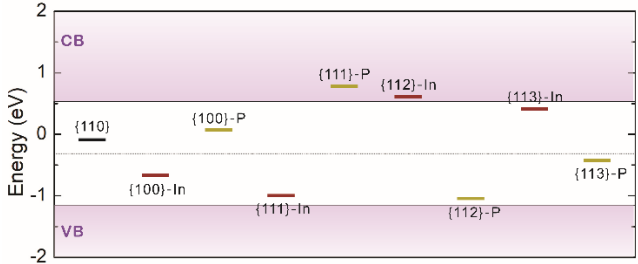

Figure 9. Integrated Fermi energy levels of the different APB structures determined with longer slabs. The black dotted line corresponds to the Fermi energy level and the purple areas corresponds to the valence bands (VB) and conduction bands (CB) of the bulk material without APBs.

In order to check the validity of these calculations and conclusions, careful convergence testing on band structures and Fermi energy level was conducted to push this study beyond the limits of convergence with longer APB slab structures (Fig. S3). Electronic bands of these longer APB slab structures were calculated based on metaGGA functionals [46-50], which are significantly less time-consuming and memory-using but show similar accuracy, as compared to the expensive HSE or GW method. Figure S4 shows the corresponding band structures and Figure 9 gives the corresponding calculated Fermi energy levels after careful calibration of the energetic positions of the deep energy levels and bulk energy states. The band structures of longer APB structures (Fig. S3) show more bands due to the addition of more atoms, and band folding effects [51]. Besides, the bands are shifted and the bulk-like CBs and VBs are closer to each other due to the weakening of the quantum confinement effect. But the most important conclusion is that the intrinsic electronic nature of the APB structures does not change: the additional states introduced by the different APB structures are still there and keep the same features (semi-conducting, metallic, with or without bandgaps) as the short APB structures.

As already discussed above, the Fermi energy levels positioning could be related in a first approximation to the APB atomic stoichiometry: P-rich APBs (like {100}-P,{111}-P,{112}-In APBs) have a relatively high Fermi energy level and In-rich APBs (such as {100}-In,{111}-In,{112}-P APBs) have a relatively low Fermi energy level (Fig.9). But on the other hand, we can find clearly that even though the {111}-P with {100}-P APBs, and {113}-P with {113}-In APBs, have the same stoichiometry, the Fermi energy level of {111}-P APB is higher than the one of the {100}-P APB, and the Fermi energy level of {112}-In is higher than the one of the {112}-P APB. This indicates there are some other determining factors for the positioning of the Fermi energy level apart from the atomic stoichiometry, which could be related to the charge density or/and bond configuration. In order to explore this, a summary of the different information including band type, III/V atomic stoichiometry, charge density, relative Fermi energy level, ratio between III-III zigzag bonds and total number of bonds, ratio between V-V zigzag bonds and total number of bonds, ratio between III-III ladder bonds and total number of bonds, ratio between V-V ladder bonds and total number of bonds for the different APBs are extracted, and shown in Table 1. As commented previously, Table 1 obviously evidences that the atomic stoichiometry cannot by itself explain the Fermi level variations. While pure group-III APBs or pure group-V APBs have indeed respectively negative and positive relative Fermi levels as expected, some large variations of the relative Fermi energy levels are indeed observed also for purely stoichiometric APBs. It is thus interesting to analyze the charge carried by APB planes. Figure 10 gives the zoomed atomic configuration of zigzag and ladder APB patterns, where we can see that atoms are in a very different situation. For the ladder case, one atom possesses only one homovalent wrong bond, and three other normal III-V bonds. For the zigzag case, one atom possesses 2 homovalent wrong bonds, and two other normal III-V bonds. As a consequence, each APB atom in a ladder configuration will lack or share 1/2 e-, while each APB atom in a zigzag configuration will lack or share 1 e-. In other words, every APB bond will give rise to an excess or a deficit of 1/2 e- in the system, the number of zigzag bonds being twice the number of ladder bonds for the same number of atoms.

Table 1: The different information of the different APBs, involving band type, atomic stoichiometry, average charge density carried by the APB, relative Fermi energy level (based on the Fermi level of bulk material without APBs, i.e. $E_{f\text{-}APB}-E_{f\text{-}bulk}$. The negative and positive Fermi energy values thus correspond to p-doped and n-doped behaviours, respectively), ratio between III-III zigzag bonds and total number of wrong bonds, ratio between V-V zigzag bonds and total number of wrong bonds, ratio between III-III ladder bonds and total number of wrong bonds, ratio between V-V ladder bonds and total number of wrong bonds.

| APB | {110} | {100}-In | {100}-P | {111}-In | {111}-P | {112}-In | {112}-P | {113}-In | {113}-P |
|---|---|---|---|---|---|---|---|---|---|
| Band type | A (Semiconducting) | B (Metallic) | B (Metallic) | C (Metallic) | C (Metallic) | A+B (Metallic) | A+B (Metallic) | A+B (Metallic) | A+B (Metallic) |
| Atomic stoichiometry (III:V) | 1:1 | pure III | pure V | pure III | pure V | 1:2 | 2:1 | 1:1 | 1:1 |
| Average charge density (e/Å$^{-2}$) | 0 | +0.0563 | -0.0560 | +0.0322 | -0.0322 | 0 | 0 | +0.0169 | -0.0170 |
| Relative Fermi level (eV) | 0.224096 | -0.35579 | 0.384996 | -0.68659 | 1.090806 | 0.918066 | -0.73215 | 0.722606 | -0.11085 |
| N(III-III zigzag)/N(total) | 0 | 1 | 0 | 0 | 0 | 1/2 | 0 | 2/3 | 0 |
| N(V-V zigzag)/N(total) | 0 | 0 | 1 | 0 | 0 | 0 | 1/2 | 0 | 2/3 |
| N(III-III ladder)/N(total) | 1/2 | 0 | 0 | 1 | 0 | 0 | 1/2 | 0 | 1/3 |
| N(V-V ladder)/N(total) | 1/2 | 0 | 0 | 0 | 1 | 1/2 | 0 | 1/3 | 0 |

Thus, for the charge density calculation, firstly, the number of the III-III and V-V APB bonds are counted (Table S1). Then we use the bond numbers multiplied by the corresponding charge per bond (V-V APB bonds donate excess electrons, with -1/2e per bond; III-III APB bonds provide holes, with +1/2e per bond) to get the total charge. Finally, the total charge is divided by 2D APB plane area (Table S1) to get the charge density. As discussed above, the relationship between the APB bond number and the APB atom number is different for the zigzag and ladder configurations, therefore we can see (in Table1) that the charge density is not directly and fully related to atomic stoichiometry for all the APBs. From the classical free electron theory [52,53], it is known that the charge density can significantly tune the Fermi energy level. The higher electron density (higher hole density) is expected to lead to higher (lower) Fermi energy level. But interestingly, from Table 1, it can be found that neither atomic stoichiometry nor charge density can provide a good and comprehensive explanation on the variation of the Fermi energy levels for all the APBs.

Thus, in the following, the specific APB bond configurations are taken into account. Based on Table 1, we can find clearly that the Fermi energy level of {111}-P with pure ladder V-V bonds is higher than the one of {100}-P with pure zigzag V-V bonds, and the Fermi energy level of {111}-In with pure ladder III-III bonds is lower than the one of {100}-In with pure zigzag III-III bonds, which indicate that the ladder APB configurations show a more pronounced effect on the relative Fermi energy levels shifting than the zigzag APB configurations (i.e. ladder APB> zigzag APB). In addition, based on the shifting values of the Fermi energy levels of {100} and {111} APBs, one can see that P-P bonds have always a larger effect on the Fermi level than In-In bonds (Fermi level is always larger "on the n side") for the same bond configuration. To summarize the results, the impact on the Fermi level shifting can be classified as: ladder P-P APB > ladder In-In APB > zigzag P-P APB > zigzag In-In APB. This surprising observation is valid for all the APBs configurations considered in this work.

To understand the fact that the ladder pattern configuration impacts the Fermi level shifting much more importantly as compared to the zigzag one, a detailed description of charges sharing is proposed. Figure.10 shows the bonding configurations of APB atoms with other APB atoms and with the surrounding bulk atoms for the zigzag (Fig.10a) and ladder (Fig.10b) APB patterns. As mentioned previously, each APB atom in zigzag pattern has two bonds connected with the other APB atoms and two bonds connected with the surrounding bulk atoms. While, for the ladder pattern, each APB atom has only one bond connected with the other APB atom and three bonds connected with the surrounding bulk atoms. Therefore, the zigzag APB configurations with stronger connection among the APB atoms and less connection with the surrounding bulk APB atoms are expected to exert stronger binding energy on the charges (Fig.10a) (i.e. the charges have lower energies). On the contrary, the ladder APB configurations with relatively weaker connection among the APB atoms and more connection with the surrounding bulk APB atoms will lead to relatively weaker binding energy on the charges (Fig.10b) (i.e. the charges have higher energies). Here we note that, normally, the band structures are based on electron energy (eV). Therefore, if the charges are electrons, then the higher energy of the electrons will lead to higher energy states, and on the contrary, if the charges correspond to holes, then the higher energy of the holes will cause lower energy states on band structure images. Overall, we can conclude that the ladder APB configurations show a much larger impact on the Fermi energy levels shifting than the zigzag APB configurations, due to the relatively lower binding energy of charges around the ladder APBs. Note that a similar trend is observed for different doping atoms in Si, that have different carriers binding energies [54-57]. Indeed, an important and simplified parallel can be made between n-doping or p-doping for the silicon semiconductor and the cases treated in this study for the APBs. So, for example, one specific APB is related to an n-like-doping or p-like-doping. However, the main differences relate to i) the dimensionality of the APB (1D or 2D) compared to classical doping with one element (0D) and ii) the fact that the APB is composed of similar atoms than the ones of the semiconducting matrix whereas different atoms are used for doping. This can be thus understood as a "self-doping" of the material.

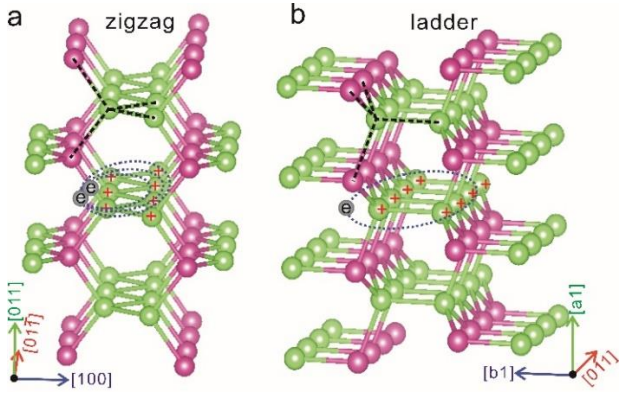

Figure 10. Schematic diagram of bound charges around the zigzag APB (a) and ladder APB (b). a. Each APB atom in zigzag pattern has two bonds connected with other APB atoms and two bonds connected with the surrounding bulk atoms (the black dotted lines), leading to relatively higher binding energy on the charges. b. Each APB atom in ladder pattern has only one bond connected with the other APB atom and three bonds connected with the surrounding bulk atoms (the black dotted lines), leading to relatively lower binding energy on the charges.

Finally, we shall discuss the influence of the different APBs on the optoelectronic properties of the III-V/Si devices. As discussed above, the different APBs introduce three new basic types of modified band structures locally in the III-V semiconductor matrix: (Type-A) new localized APB states on the top of VBs, reducing bulk-like bandgap but still keeping semiconductor nature (e.g. {110} APBs); (Type-B) new metallic states localized around APBs connecting CBs and VBs, closing the bulk-like bandgap and turning it to metallic nature (e.g. {100} APBs); (Type-C) new localized metallic APB states on the top of VBs and bottom of CBs, reducing the bulk-like bandgap and turning it to metallic nature (e.g. {111} APBs) (as shown in Fig.11).

From the charge carriers transport point of view, firstly, the Type-B and Type-C band structures introducing metallic states, the corresponding APBs (such as {100}, {111}, {112}, {113} APBs) are thus expected to behave as efficient transport channels, thus explaining the electrical shortcuts observed in devices with APBs [19]. Besides, even though the Type-A band structure still keep a semiconducting nature, strong electron-phonon coupling could also benefit to carriers transport [2]. Therefore, overall, relatively good transport properties are expected for the different APBs, and especially the metallic ones, which provides a comprehensive frame to interpret the recent results obtained on GaP/Si samples by Conductive-Atomic Force Microscopy (C-AFM) measurements [11].

On the other hand, from the optical properties point-of-view, electronic band gaps of materials provide paths for electron-hole generation through light absorption, or electron-hole recombination through light emission (Fig.11a). The Type-A and Type-C band structures with a reduced but opened bandgap can still enable light absorption and light emission (Fig.11b,d). We note here that the Type-A corresponds to a very standard semiconductor configuration, while in the case of type-C the efficiency of light emission or absorption processes strongly depends on Fermi level positioning relative to the bands, which would require experimental investigations that are beyond the scope of this work. We also note here that there exist different APB configurations that can contribute to light emission processes. This result could explain why significant inhomogeneous broadening was observed on both APBs-related PL peaks as well as on phonons modes observed in Raman spectroscopy in previous works [2]. On the contrary, the Type-B band structure without bandgap isn't expected to give rise to light emission due to the direct non-radiative recombination of excited electrons and holes through the metallic states, as shown in Fig.11c.

Finally, from the device point-of-view, the different APBs that have different Fermi energy levels are thus expected to behave as acceptors or donors. When the APBs couple together with the bulk material, charge redistribution occurs and a built-in electric field can form [11]. This built-in electric field can work as a driving force to separate photogenerated carriers. Therefore, the APBs enable efficient carrier separation and collection [11]. The present work strengthens the idea that bidomain epitaxial III-V/Si materials are indeed composed of numerous lateral micro- or nano- p-n junctions, whose electric field has a significant in-plane component. Overall, carrier localization effects available in APBs having Type-A band structures could be very interesting for management of light emission at the nanoscale, and APBs having Type-B or Type-C band structures seem more interesting for carrier extraction in light harvesting devices (e.g. for solar cells or solar water splitting devices).

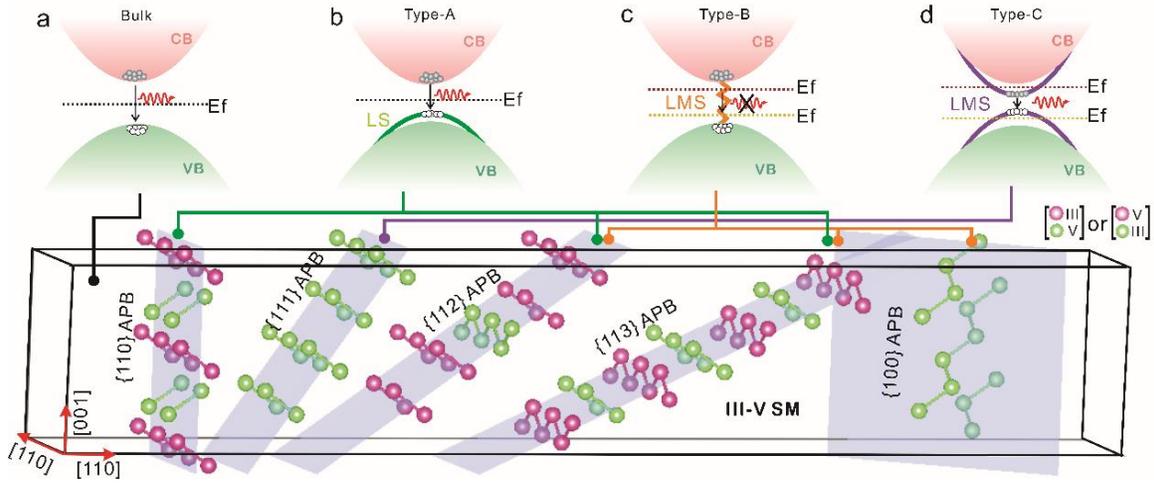

Figure 11. Schematic plot of the relationship between the different APBs configurations with the three basic APB band structures (b-d), as compared to the bulk band structure (a). (b) Type-A gapped semiconducting APB band structure corresponding to the stoichiometric APB with a ladder pattern wrong bond configuration (e.g. {110} APB). The green line in (b) represents the APB localized state (LS). (c) Type-B non-gapped metallic APB band structure corresponding to the APBs with zigzag pattern wrong bond configuration (e.g. {100} APB). (d) Type-C gapped metallic APB band structure corresponding to the non-stoichiometric APB with ladder pattern wrong bond configuration (e.g. {111} APB). The orange line in (c) and purple lines in (d) represent the APB localized metallic states (LMS). The band structures of {112} and {113} APBs are combinations of Type-A and Type-B band structures due to the mixture of ladder and zigzag wrong bonds. The Fermi energy levels are marked by dotted lines. For the Type-B and -C band structures, due to different stoichiometry or bond configuration, the Fermi energy level can be relatively high (dark red dotted line) or low (dark yellow dotted line). In order to show the effect of APBs on the optical properties of III-V/Si materials, the recombination of excited electrons and holes are marked on the band structures.

## V. CONCLUSION

In conclusion, we comprehensively studied and analyzed the atomic structures and electronic band structures of different antiphase boundaries (APBs), including {110}, {100}, {111}, {112} and {113} APBs based on first-principle calculations. We demonstrate that APBs can be classified following three different band structures: gapped semiconducting band structure, gapped metallic band structure and non-gapped metallic band structure. We also show how these properties are closely linked to the stoichiometry and bond configuration of the considered APBs. More specifically, we point out that the atoms involved in the wrong bonds of APBs can be organized following a ladder or a zigzag pattern. We demonstrate that the ladder pattern wrong bond configuration introduces localized states at the top of VBs and/or at the bottom of CBs while the zigzag pattern wrong bond configuration introduces states localized around the APB connecting the bulk-like VBs and CBs and closing the electronic band gap. We also conclude that the ladder APB configurations show more significant impact on the Fermi energy levels than the zigzag APB configurations, due to the relatively lower binding energy of the charges around ladder APBs. A simplified parallel can be made between n/p-doping for the variation of the energy Fermi level. Moreover, here the main difference relates to dimensionality (here 1D or 2D) and the doping is not inherent to a new element included in the semiconductor but the signature of the APB. Finally, we show that these findings lead to consider III-V/Si bidomain epitaxial materials as composed of micro- and nano-p-n junctions, and discuss prospectively these materials for use in photonic or energy harvesting devices.

This research was supported by the French National Research Agency PIANIST Project (Grant no. ANR-21-CE09-0020), NUAGES Project (Grant no. ANR-21-CE24-0006). DFT calculations were performed at Institut FOTON and the work was granted access to the HPC resources of TGCC/CINES/IDRIS under the allocation 2020-2022 A0080911434, A0100911434 and A0120911434 made by GENCI.